\documentclass[a4paper]{article}

\usepackage{INTERSPEECH2019}

\usepackage{subcaption}
\usepackage{hyperref}
\usepackage[T1]{fontenc}
\newcommand\blfootnote[1]{%
  \begingroup
  \renewcommand\thefootnote{}\footnote{#1}%
  \addtocounter{footnote}{-1}%
  \endgroup
}

\title{Detection of Glottal Closure Instants from Raw Speech using Convolutional Neural Networks}
\name{Mohit Goyal*\thanks{* Equal Contribution}, Varun Srivastava*, Prathosh AP}
\address{
  Indian Institute of Technology, Delhi}
\email{\{goyal.mohit999, varunsrivastava.v, prathoshap\}@gmail.com}

\begin{document}
\maketitle
\begin{abstract}Glottal Closure Instants (GCIs) correspond to the temporal locations
  of significant excitation to the vocal tract occurring during the
  production of voiced speech. GCI detection from speech signals
  is a well-studied problem given its importance in speech processing.
  Most of the existing approaches for GCI detection adopt a two-stage
  approach (i) Transformation of speech signal into a representative
  signal where GCIs are localized better, (ii) extraction of GCIs using
  the representative signal obtained in first stage. The former stage
  is accomplished using signal processing techniques based on the principles
  of speech production and the latter with heuristic-algorithms such
  as dynamic-programming and peak-picking. These methods are thus task-specific
  and rely on the methods used for representative signal extraction.
  However in this paper, we formulate the GCI detection problem from
  a representation learning perspective where appropriate representation
  is implicitly learned from the raw-speech data samples. Specifically,
  GCI detection is cast as a supervised multi-task learning problem
  solved using a deep convolutional neural network
  jointly optimizing a classification and regression cost. The learning
  capability is demonstrated with several experiments
  on standard datasets. The results compare well with the state-of-
  the-art algorithms while performing better in the case of presence
  of real-world non-stationary noise.
\end{abstract}

\noindent\textbf{Index Terms}: GCI detection, epoch extraction, dilated convolutional neural networks, multi-task learning.

\blfootnote{The code used for implementation including architecture and hyperparameter settings is available at \url{https://github.com/VarunSrivastavaIITD/DCNN}.}
\section{Introduction}

\subsection{Background and Previous work}

Production of voiced speech is accompanied with sustained oscillations
of the vocal folds \cite{fant1971acoustic} resulting in a quasi-periodic
flow of air-pulses which constitutes the excitation signal to the
vocal tract \cite{story2002overview}. The instant of significant
excitation (within each period) is termed as Epoch which coincides
with instant of closure of the glottis \cite{TVA79}. The problem
of detecting the precise locations of such Glottal Closure Instants
(GCIs) from speech signal has been studied for decades given its importance
in several speech processing tasks \cite{TVABY75,DYPSA2,yaga1,ZFR,SEDREAMS,6562799,prathosh2016cumulative,khanagha2014detection,gcireview}.
Most of the successful GCI detectors adopt a two-stage approach -
(i) Obtaining an intermediate representation from the speech signal, which explicitly manifests GCIs as discontinuities, impulses, extremas or as other perceptual events and (ii) detecting precise temporal location of glottal closures using custom-made algorithms. The former stage is
based on the observation that GCIs are not comprehensible in the raw
speech-signal domain but exhibit better localization in some other
domain (Figure 1 in \cite{6562799}). Many algorithms rely on the
source-filter model for speech production \cite{fant1971acoustic}
and use signal-processing techniques to estimate a correlate of the
source-signal, in which GCIs are better manifested. For instance,
\cite{DYPSA2,yaga1,prathosh2016cumulative,koutrouvelis2016fast} choose
either linear-prediction residual or glottal flow derivative as the
representative signal. Other class of algorithms do not explicitly
make any model assumption for speech production rather indirectly
use the properties of excitation signal (such as its impulsive nature)
and estimate appropriate representations (E.g., zero-frequency filtered
signal \cite{ZFR}, mean-based signal \cite{SEDREAMS}, wavelet-decompositions
\cite{d2011glottal}, singularity exponents \cite{khanagha2014detection}).
During the second stage, aforementioned algorithms employ several
heuristics to extract (or refine) the GCIs. These include dynamic
programming \cite{DYPSA2,yaga1}, peak-picking \cite{6562799,6080715}
and optimization with regularity constraints \cite{khanagha2014detection}.
All these methods perform reasonably well albeit they depend largely
on their choice of representative signals.

\subsection{Context and Scope}

With the recent advances made in the area of data-driven representation
learning \cite{bengio2013representation}, it is possible to operate
directly in the signal space and let the learning machinery obtain
the appropriate representations given any task and data. This approach has
found tremendous success in multiple domains with image, text and
audio data \cite{goodfellow2016deep}. Specifically, convolutional
neural networks (CNN) have found their utility in a range of speech
processing tasks such as phoneme recognition \cite{doss1,doss6},
feature/front-end learning for LVCSR \cite{doss3,doss2,abdel2014convolutional},
voice-activity detection \cite{doss4}, spoofing detection \cite{dinkel2017end},
emotion recognition \cite{trigeorgis2016adieu,zhang2017speech} and
speaker identification \cite{doss5} . The underlying theme in all
these methods is to directly operate on the raw speech signal and
let the CNN learn the abstract representations resulting in superior performance as compared to hand-crafted feature engineering.

Now, data driven algorithms have already been employed on the problem of GCI detection with a significant improvement in localization error and detection accuracy. Several standard machine learning methods such as extremely randomized trees, SVM, k-nearest neighbours and multilayer perceptron over hand engineered features obtained from speech signal have been used for GCI detection \cite{interspeechML}. CNNs are also shown to be much more proficient in extracting epochs than non-data-driven methods on pathological acoustic speech acquired from vocally impaired patients \cite{NeurIPSCNN}. Motivated by the above observations, in this paper, we approach the problem of end-to-end GCI detection from a deep-learning perspective. The key difference in the present formulation as compared to the previous
works which uses CNNs for various speech processing tasks is that,
by-and-large the scope of previous works is classification of an utterance
or segment of speech into classes (phonemes, emotions, speakers) while
in the proposed work the interest is in detecting a temporal event
(GCI) in the signal. This objective is met by formulating the problem
in a novel joint classification-regression framework wherein a temporal
event (GCI in this case) is simultaneously detected and localized
in a frame. Various experiments are performed
on multiple datasets comparing with four state-of-the-art algorithms
to demonstrate effectiveness of the proposed method through improved detection and localization metrics. Previous work on utilizing CNNs for GCI localization \cite{yang2018detection} operates on low pass filtered speech and relies on the assumption that every GCI coincides with a negative peak in the speech waveform. However, in this work, we pose GCI extraction as a general temporal event detection problem relaxing both the constraints.

\section{Methodology}

\subsection{Problem formulation and Data generation}

In this work, the problem of GCI detection is formulated using a block-processing
approach. It is known that GCI is a temporal event that occurs atmost
once in every pitch-period of the voiced speech that could range between
2 milli-seconds (ms) to 20 ms \cite{titze1998principles}. Based on this observation,
we define a detection speech window ($w_{d}$) as any speech frame
of length equal to 2 ms (minimum possible pitch period). Therefore, every detection window can have atmost one GCI within
it. Given a $w_{d}$, the primary task is to detect whether or not
there is GCI within it, which is formulated as a binary classification
problem. Having detected a GCI within a $w_{d}$, the next step is
to localize the GCI which is cast as a regression task, estimating
the distance between the onset of $w_{d}$ and the location of GCI.
Since a 2 ms window comprises too less input features (32 samples at 16kHz) for meaningful feature learning via a deep CNN, a symmetric context of 1.5 ms at either sides of the detection
window is considered to generate the input frame $w_{i}$. Hence in
summary, every input data-sample ($w_{i}$) will be of length 5 ms,
with the classification and regression for GCI detection and localization
being carried simultaneously. Finally, all input data-frames of size $w_i$ are generated with a shift of one
sample between successive input frames. In principle, heuristics can be designed to reject data-frames (with no associated ground-truth GCI) based on frequency components, amplitude or other signal attributes. However, we refrain from performing any pre-processing/filtering on the input speech to solve the problem in an end to end fashion. Therefore, each possible data-frame is considered for the purpose of epoch extraction.

\subsection{Dilated CNNs and Multi-task learning}

Recently, Deep Dilated Convolutional Neural Networks have been shown
to learn useful representations for several speech processing tasks \cite{yu2015multi}\cite{van2016wavenet}. Since, a dilated
convolution is mathematically equivalent to regular convolution with
a kernel with zeros inserted in between, stacked dilated convolutions
allow an exponential increase in context with a linear increase in parameters. The resulting reduction in parameters provides a regularization effect to the CNN improving its generalization on test data. The proposed neural architecture comprises of  convolutional layers that function as feature extractors, so that a \emph{shared} representation is obtained for a speech window. Max pooling is used after every convolutional layer with a kernel size and stride of 2 while doubling the dilation at every convolutional layer to increase the corresponding receptive field at each layer. Scaled Exponential Linear Units (SELU) \cite{klambauer2017self} activation functions are used between convolutional layers in order to normalize the activations. SELU ensures the benefits of explicit normalization (E.g.,  batch-normalization \cite{ioffe2015batch})  while being identical implementation-wise during train and test phase, unlike techniques such as batch normalization.

Since both the classification and regression tasks are to be performed
on the same input frame ($w_{i}$) the features are jointly learned
which is the core idea behind multitask learning \cite{caruana1998multitask}. Thus, an input frame $w_{i}$ is first fed to the CNN to extract the feature vector that goes into two identical subnetworks performing classification and regression. A sigmoid activation is used at the output of the classification branch and a Hard-tanh function is used for regression (instead of a linear activation to suppress out-of-window predictions).

The task is to output the probability of presence of a GCI $y_{c}\; (0 \leq y_c \leq 1)$,
as well as its location $y_{r}$ ($0\le y_{r}<w_{d}$), within each
detection window. Ideally, for a window with a GCI, true $y_{c}=1$ and
true $y_{r}=d$ (where $d$ is the  distance between the onset of the window
and location of the GCI within the window) and for a window without
a GCI, true $y_{c}=0$, and true $y_{r}$ can be anything and is immaterial (Refer eq.\eqref{eq: loss fn}). Since
a single network predicts both the probability of occurrence of the
GCI and location of the GCI, the loss function consists of two terms,
the classification error which is standard cross entropy loss and
the regression error which is the mean squared loss between the actual
and the predicted location of the GCI within a window. Mathematically,
\begin{align}\label{eq: loss fn}
  \mathcal{L} & = \frac{c_{1}}{N}\sum_{j=1}^{N}(t_c^j log(y_c^j) + (1-t_c^j) log(1-y_c^j)) \nonumber \\ &+ \frac{c_2}{\sum_{k=1}^{N}t_c^k}\sum_{j=1}^{N} t^j_c*(t_r^j-y_r^j)^2\end{align}

where $t_c^j$ is $1$ if the $j^{th}$ sample window has a GCI, and is $0$ otherwise. $t_r^j$ is the actual location of the GCI within the $j^{th}$ window, $N$ is the number of datapoints within a training batch. $c_{1}$ and $c_{2}$ are the weights of the classification and regression loss respectively, found empirically using grid search. Note that the regression loss is divided by the number of true GCIs in a batch (instead of total number of datapoints) to account for the fact that there are fewer samples for the regression block in every batch compared to the classification block.

\subsection{Inference and Weighted Histogram Clustering (WHC)}

\begin{figure}[!htbp]
  \centering
  \includegraphics[width=0.47\textwidth]{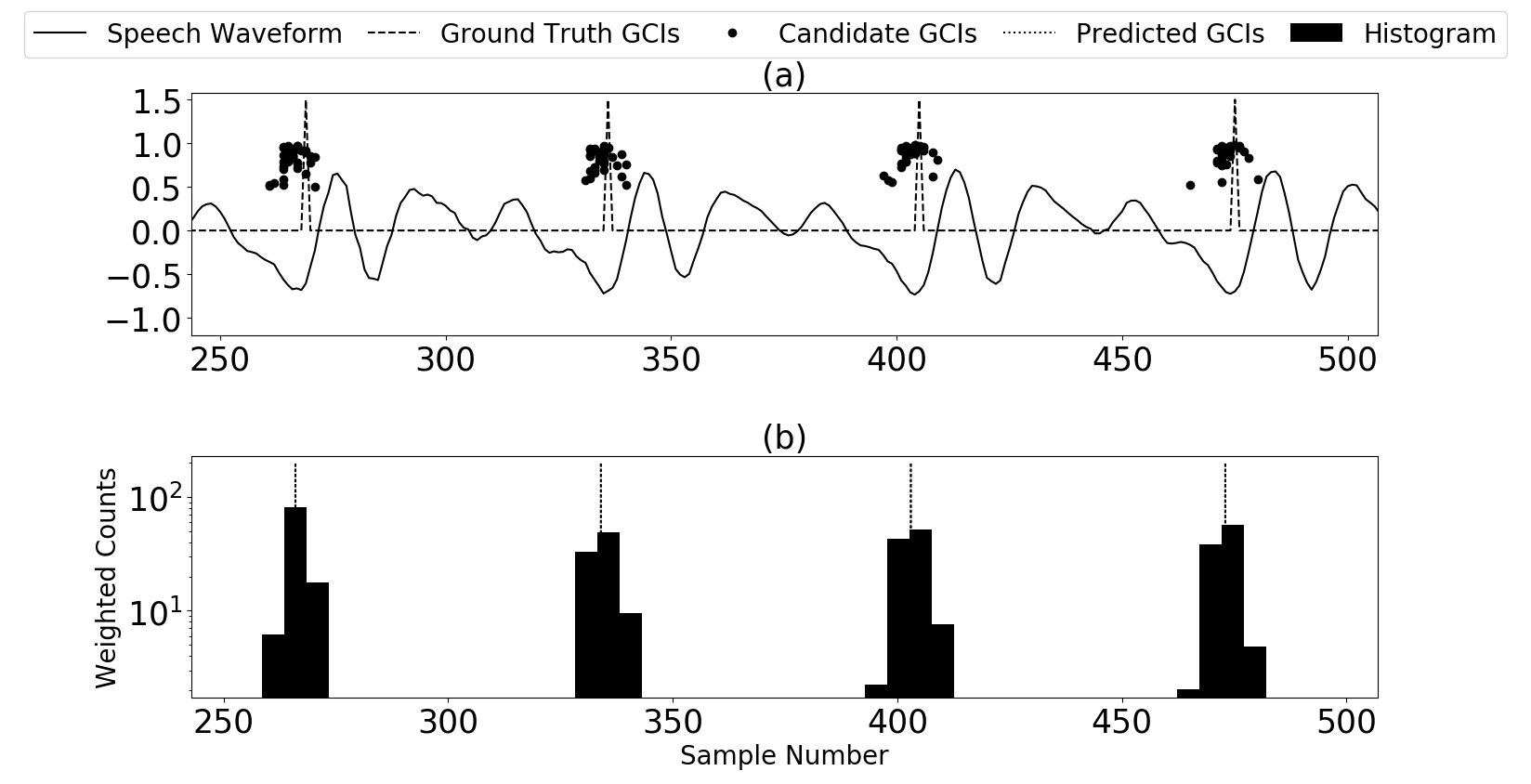}
  \caption{Illustration for WHC (a) a
    segment of speech with four ground truth GCIs along with the corresponding candidate GCIs (height represents confidence in detection), (b) weighted histograms shown along with the means of the local-groups (dotted lines) which form the final predictions.}
  \label{fig: cluster}
\end{figure}

\begin{figure}[!htbp]
  \centering
  \includegraphics[width=0.38\textwidth]{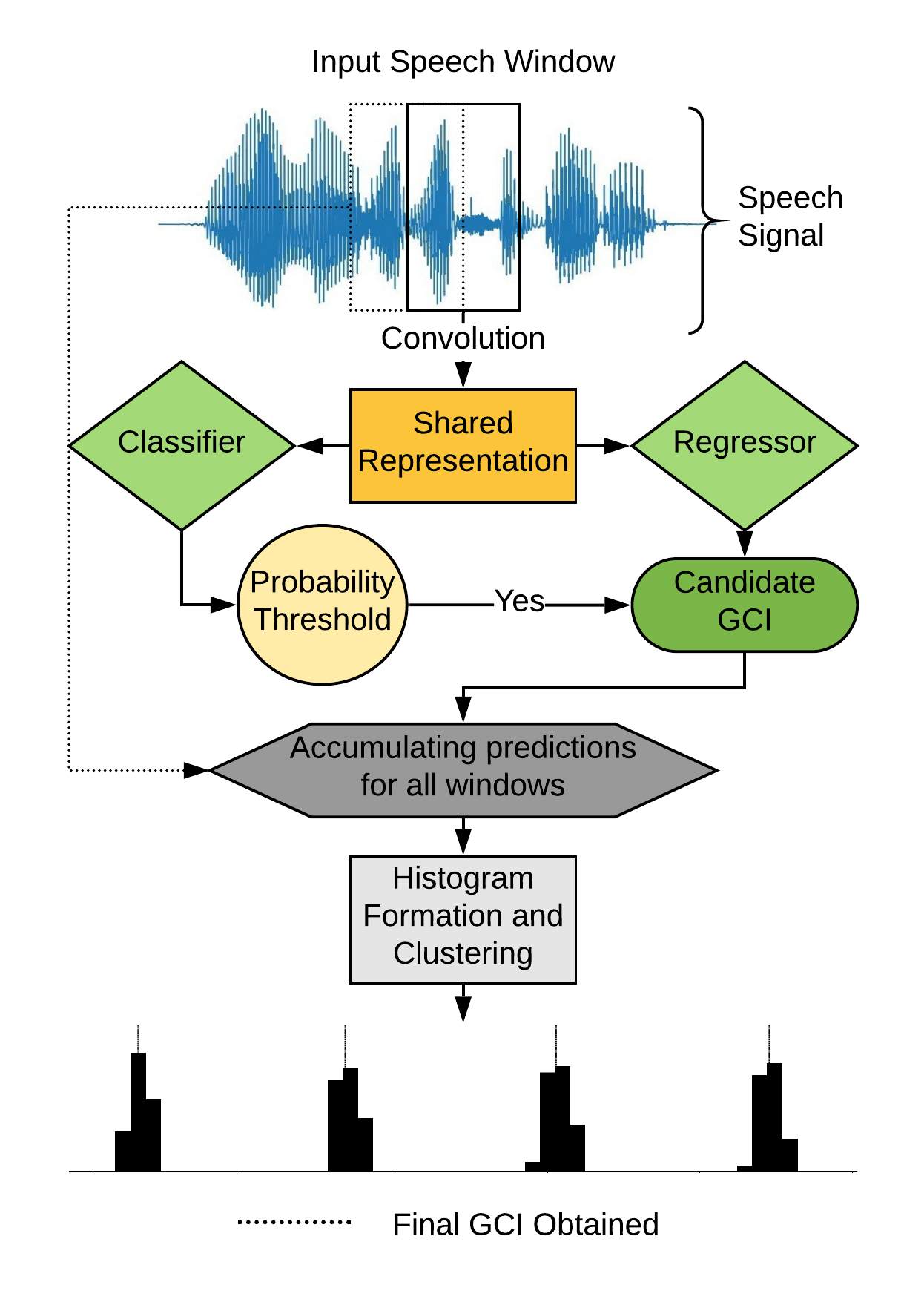}
  \caption{Proposed Algorithm: Trained CNN outputs a probability and location for each window. The windows satisfying probability threshold are considered as GCI candidates. WHC algorithm filters though them giving out final GCIs.}
  \label{fig: Algo}
\end{figure}

In the present formulation, every input data-frame/window contains
atmost one GCI. Since one ground truth GCI can belong to multiple data-frames, possibly, there can be duplicate predictions over successive input windows resulting in false detections.
Hence, there is a need for multiple predictions to be  merged together into a single predicted GCI. Given a speech signal and its associated candidate GCIs, the first step is to construct a weighted histogram of the candidate GCIs with a bin size of $B$ (hyper-parameter). The confidence associated with each candidate for the presence of GCI is used to calculate the weight for the histogram. Since occurence of glottal closure is a quasi-periodic event, the histogram thus obtained will contain repeating local-groups of contiguous bins with non-zero values (Fig. \ref{fig: cluster}(b)). The final GCIs are hypothesized to be the means (measure of central tendency) of local-group thus (dotted lines in Fig. \ref{fig: cluster}(b)) obtained. These means tend to be at the locations with large number of high-probability candidate GCIs which is desirable. The bin-size $B$ is set at 5 samples through-out this work albeit there is a trade-off between erroneous detections (higher the $B$ lesser are the false insertions) and resolution of predicted GCIs (lower the $B$, better is the localization of detected GCIs). The entire procedure for clustering is illustrated in Figure \ref{fig: cluster}. The complete algorithm is depicted in Figure \ref{fig: Algo}.

% \begin{figure}[!htbp]
%   \centering
%   \includegraphics[trim={20 18 18 15},clip,width=0.47\textwidth]{./fig_4_22_44_48.eps}
%   \caption{tsne}
%   \label{fig: tsne_class}
% \end{figure}

% \begin{figure}[!htbp]
%   \centering
%   \includegraphics[trim={20 18 18 15},clip,width=0.47\textwidth]{./fig_4_22_53_39.eps}
%   \caption{tsne}
%   \label{fig: tsne_regress}
% \end{figure}

\section{Experiments and Results}

\begin{figure*}[!htbp]
  \centering
  \includegraphics[width=\textwidth]{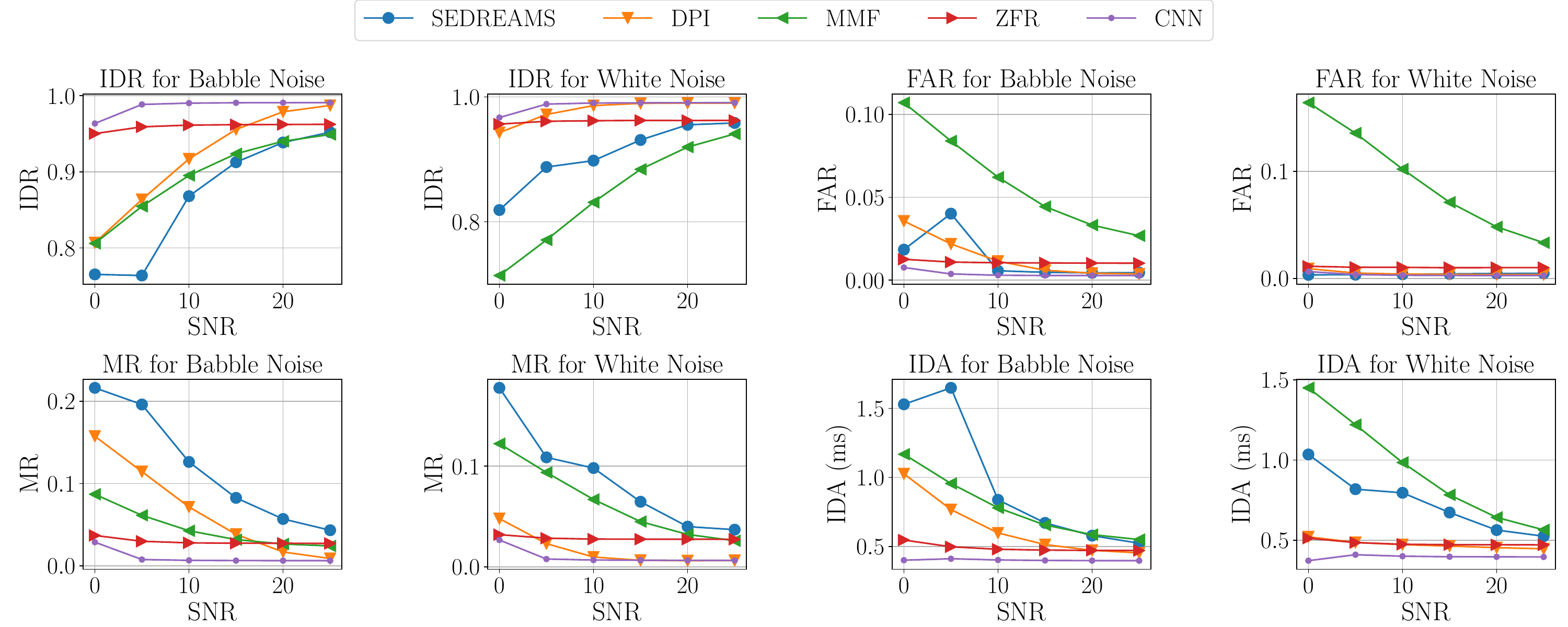}
  \caption{Comparison of five algorithms for different noise levels. It can be seen that the proposed algorithm (CNN) is comparable to the best of the state-of-the-art methods for all cases considered. }
  \label{fig: comparison}
\end{figure*}

\subsection{Experimental details}

The proposed GCI detection algorithm is evaluated on
datasets containing simultaneous recordings of speech and Electroglottography
(EGG) signals. The negative peaks obtained from differentiated
EGG (dEGG) signals form the ground truth for GCI locations. In this paper, we use speech signals from the CMU Arctic \cite{kominek2004cmu} dataset for our experiments. Standard performance metrics namely, Identification
Rate (IDR - \% of correct detections, higher the better), Miss Rate
(MR - \% of missed detections, lower the better), False Alarm Rate
(FAR - \% of false insertions, lower the better) and Identification
Accuracy (IDA - standard deviation of distance between the true and
predicted GCIs, lower the better) are employed for evaluation. (A detailed description for metrics may be found in Figure 2 of \cite{DYPSA2}).
% and elsewhere \cite{gcireview,6562799}.
There are in total of 3 speakers in the CMU dataset---JMK (Canadian Male), BDL (US Male), SLT (US Female). The CNN model is implemented using PyTorch \cite{team2017pytorch} using the ADAMAX optimizer \cite{kingma2014adam}
with standard parameter settings.
Model parameters were initialized as per the scheme outlined in \cite{klambauer2017self}. Detection
experiments were carried out on clean speech as well as speech corrupted with additive synthetic white noise and real-world babble noise (background multi-speaker chatter obtained from \cite{nd}) with SNRs ranging from 0 to 25 dB in steps of 5 dB. For a given SNR, both the training and testing are carried on the corresponding corrupted speech at the same SNR. The results of the proposed algorithm (CNN) are compared with four state-of-the-art algorithms, namely, Zero-frequency resonator \cite{ZFR}, Speech Event Detection using the Residual Excitation And a Mean-based Signal (SEDREAMS) \cite{SEDREAMS}, Dynamic Plosion Index (DPI) \cite{6562799} and Micro-canonical Multi-scale Formalism (MMF) \cite{khanagha2014detection}, by evaluating all four on the same test dataset as that of CNN. Training and testing is performed on CMU corpus with speaker-overlap between the training and test data. Both training and test data contain utterances (non-overlapping across training and test) from all speakers. We also compare the results of extremely randomized trees (ERT) (introduced in \cite{interspeechML} for GCI detection) with the proposed algorithm. Since ERT is originally proposed only for clean speech, the comparison is limited to noise free speech.
% {\color{purple}
% (b) Cross-dataset
% experiments: Training on CMU and testing on APLAWD and vice-versa,
% (c) Cross-speaker experiments: Training and testing on non-overlapping
% set of speakers, (d) Cross-SNR experiments: Training on a given SNR
% and testing on other SNRs. While experiment (a) evaluates the detection
% capabilities of the proposed method (hence this is used for comparison),
% the rest of experiments will validate its generalization capabilities.}
In all experiments a random subset of 10\% of data is considered for
training and rest for testing. Since 90\% of the data is used for
testing, no cross-validation is employed.

\subsection{Results and Discussion}

\begin{figure}[!htbp]
  \centering
  \begin{subfigure}[b]{0.45\textwidth}
    \centering
    \includegraphics[width=\textwidth]{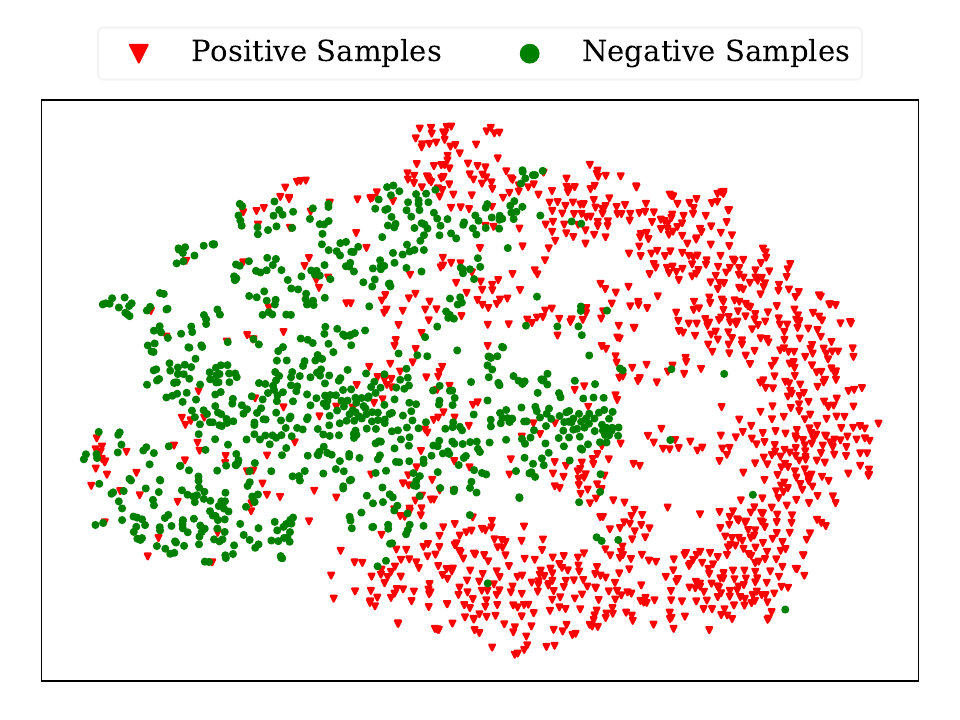}
    \caption{}
    \label{fig: tsneclass}
  \end{subfigure}
  \begin{subfigure}[b]{0.5\textwidth}
    \includegraphics[width=\textwidth]{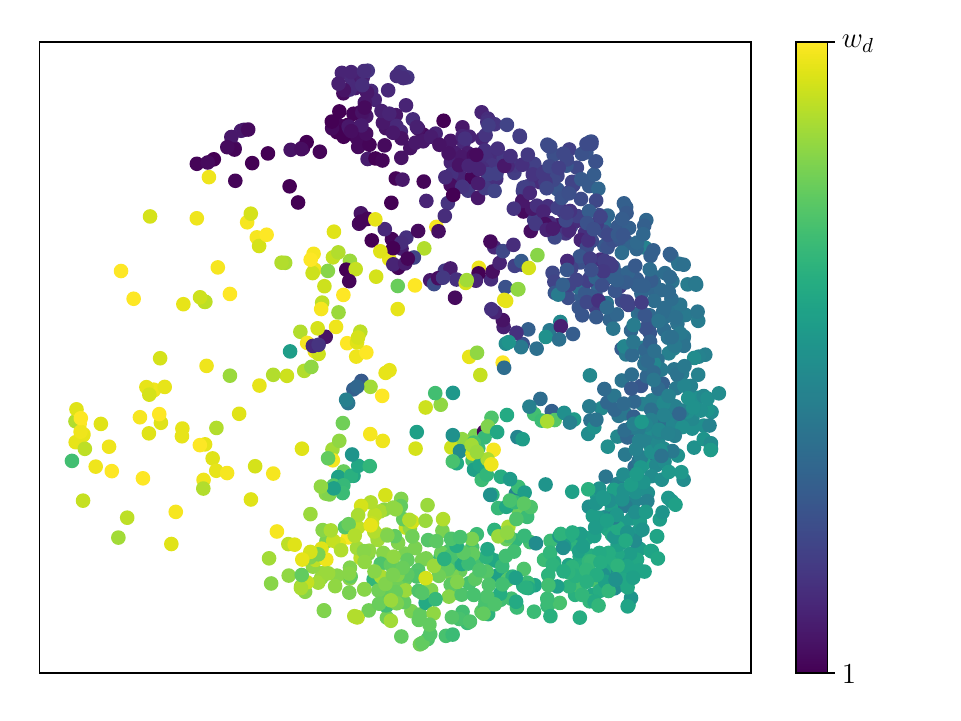}
    \caption{}
    \label{fig: tsneregress}
  \end{subfigure}
  \caption{Visualizing shared representation (extracted from CNN) with the help of t-SNE using (a) classification labels. (b) regression values corresponding to color gradient.}
  \label{fig: TSNE}
\end{figure}
Table \ref{tab: resulttab} describes the results of six algorithms on the CMU ARCTIC on clean speech. It can be seen that CNN has the highest detection rate (99.3 \% and 95.5 \%) on both the datasets. The IDA of the CNN method is also the least for both the datasets.

% \begin{table}[tbh]

% \centering{}%
% \begin{tabular}{|c|c|c|c|c|}
% \hline 
% Metrics & IDR (\%) & MR (\%) & FAR (\%) & IDA (ms)\tabularnewline
% \hline 
% \hline 
% DCNN & \textbf{99.3, 95.5}  & 0.3\textbf{, 2.2} & \textbf{0.4, }2.3 & \textbf{0.2, 0.3}\tabularnewline
% \hline 
% ZFR & 96.24, 95.2 & \textbf{2.7} 3.7 & 1.0, 1.0 & 0.47, 0.4\tabularnewline
% \hline 
% SEDREAMS & 92.6, 88.9 & 7.4, 7.5 & 0.4, 3.5 & 0.8, 0.5\tabularnewline
% \hline 
% DPI & 98.9, 94.8 & 0.4, 3.8 & 0.3, 1.3 & 0.4\textbf{, }0.5\tabularnewline
% \hline 
% MMF & 96.0, 90.5 & 2.1, 5.0 & 1.7, 4.4 & 0.5, 0.6\tabularnewline
% \hline 
% \end{tabular}
% \caption{Results on clean speech from CMU Arctic (first column in each row)
% and APLAWD (second column in each row) on five GCI detectors. {\color{red} Add row for ERT-P3}}
% \end{table}

\begin{table}[tbh]
  \centering{}%
  \begin{tabular}{|c|c|c|c|c|}
    \hline
    Metrics  & IDR(\%)       & MR(\%)       & FAR(\%)      & IDA(ms)\tabularnewline
    \hline
    \hline
    DCNN     & \textbf{99.3} & \textbf{0.3} & \textbf{0.4} & \textbf{0.2}\tabularnewline
    \hline
    ZFR      & 96.24         & 2.7          & 1.0          & 0.47\tabularnewline
    \hline
    SEDREAMS & 92.6          & 7.4          & 0.4          & 0.8\tabularnewline
    \hline
    DPI      & 98.9          & 0.4          & 0.3          & 0.4\tabularnewline
    \hline
    MMF      & 96.0          & 2.1          & 1.7          & 0.5\tabularnewline
    \hline
    ERT      & 93.57         & 2.16         & 4.26         & 0.28\tabularnewline
    \hline
  \end{tabular}
  \caption{Results on clean speech from CMU Arctic (BDL, JMK, SLT averaged) on six GCI detectors.}
  \label{tab: resulttab}
\end{table}

A similar trend is observed even with noisy speech as can be observed
in Figure \ref{fig: comparison}. It is seen that CNN and ZFR are very robust to noise
with the IDR being above 90\% even at 0 dB babble noise. Further,
the IDA of the CNN algorithm is consistently lowest for all cases considered.
% \begin{table}[tbh]

% \begin{tabular}{|c|c|c|}
% \hline 
% Experiments & IDR(\%) & IDA(ms)\tabularnewline
% \hline 
% \hline 
% Train CMU - Test APLAWD & 93.2 & 0.28\tabularnewline
% \hline 
% Train APLAWD - Test CMU & 96.3 & 0.4\tabularnewline
% \hline 
% Train Male Spks. - Test Female Spk. & 98.8 & 0.29\tabularnewline
% \hline 
% Train Female Spk. - Test Male Spks. & 97.5 & 0.35\tabularnewline
% \hline 
% Train 0 dB - Test 10 dB & 97.9 & 0.66\tabularnewline
% \hline 
% Train 0 dB - Test 20 dB & 97.9 & 0.68\tabularnewline
% \hline 
% Train 0 dB - Test clean speech & 96.55 & 0.58\tabularnewline
% \hline 
% \end{tabular}
% \caption{Results of the proposed algorithm on several generalization experiments.}
% \centering{}
% \end{table}

% Table II describes the outcomes on the experiments (b), (c) and (d)
% that are designed to ascertain the generalization abilities of the
% proposed algorithm across different speaker, dataset and SNR settings.
% For the case of cross-dataset studies (Noe that CMU has three speakers
% while APLAWD has ten), the IDR reduces by 2-3\% albeit IDA remains
% intact. To further confirm the robustness to pitch-period, we trained
% a model on only male speakers and tested on the female speaker of
% the CMU and vice-versa. It was the observed that there was no or slight
% reduction in IDR and IDA, compared to the models that were trained
% on all speakers. In the third experiment, a model that was trained
% on 0 dB Babble noise was tested on speech with different SNRs and
% it was observed that there is no change in the model performance.

It is noteworthy that in all these experiments, the model parameters (detection probability threshold: 0.5 and bin-size: 5) were kept the same in-spite of which the models generalize well across datasets, recording settings, speakers and noise-levels. However, in practical settings, one can tweak the model hyper-parameters to best suit the data. Given that the proposed method operates on raw speech and fully data-driven, the aforementioned performance is significant. All the results confirm the learning abilities of the proposed architecture. The performance of DCNN might be ascribed to the following factors: (i) problem formulation that enforces presence of multiple
detections per GCI leading to robustness, (ii) a multi-task approach to learn a shared feature representation, (iii) use of weighted histogram for clustering.

% We also provide a way to observe whether the features extracted by the CNN can act as valid representations for both the tasks. This is necessary and justifies the multi-task learning approach opted in this paper. We apply t-Distributed Stochastic Neighbor Embedding (t-SNE) technique for dimensionality reduction on the features extracted by the convolutional layers.
We also visualize the representations learnt by the shared convolutional layers using t-Distributed Stochastic Neighbor Embedding (t-SNE) \cite{maaten2008visualizing} technique for dimensionality reduction.
It can be seen that the data-points are easily segregated on the basis of class (i.e. whether a given window contains a GCI) as shown in Figure \ref{fig: tsneclass} showing that the learnt representation is conducive for high classification accuracy. Simultaneously, we can also observe in Figure \ref{fig: tsneregress} that the positive samples corresponding to windows containing GCIs follow a well defined manifold according to different positions of the GCI within a window. Hence, these representations provide a justification for using a multi-task learning paradigm which learns efficient representations suitable for multiple downstream tasks (i.e. classification and regression).

\section{Conclusion}

In this paper, a data-driven method for GCI detection from raw-speech using a multi-task supervised learning approach is proposed. A CNN is employed for learning a shared representation for two tasks simultaneously followed by a distribution estimation method for inference and clustering. Several experiments were conducted  to compare the performance  with  state-of-the-art algorithms on data-sets comprising multiple speakers to demonstrate the efficacy of the proposed approach. Given the representation and generalization abilities of the proposed approach, we believe that a similar methodology could be adopted for detecting multiple landmarks occurring in speech and  general time-series data, which could provide directions for future work.

\bibliographystyle{IEEEtran}

\bibliography{ref,epoch_reference}

% \begin{thebibliography}{9}
% \bibitem[1]{Davis80-COP}
%   S.\ B.\ Davis and P.\ Mermelstein,
%   ``Comparison of parametric representation for monosyllabic word recognition in continuously spoken sentences,''
%   \textit{IEEE Transactions on Acoustics, Speech and Signal Processing}, vol.~28, no.~4, pp.~357--366, 1980.
% \bibitem[2]{Rabiner89-ATO}
%   L.\ R.\ Rabiner,
%   ``A tutorial on hidden Markov models and selected applications in speech recognition,''
%   \textit{Proceedings of the IEEE}, vol.~77, no.~2, pp.~257-286, 1989.
% \bibitem[3]{Hastie09-TEO}
%   T.\ Hastie, R.\ Tibshirani, and J.\ Friedman,
%   \textit{The Elements of Statistical Learning -- Data Mining, Inference, and Prediction}.
%   New York: Springer, 2009.
% \bibitem[4]{YourName17-XXX}
%   F.\ Lastname1, F.\ Lastname2, and F.\ Lastname3,
%   ``Title of your INTERSPEECH 2019 publication,''
%   in \textit{Interspeech 2019 -- 20\textsuperscript{th} Annual Conference of the International Speech Communication Association, September 15-19, Graz, Austria, Proceedings, Proceedings}, 2019, pp.~100--104.
% \end{thebibliography}

\end{document}